\documentclass[twocolumn,prl]{revtex4-1}
\usepackage[english]{babel}
\usepackage{graphicx}
\usepackage[applemac]{inputenc}
\usepackage{textcomp}
\usepackage{epsfig}
\begin{document}
\title{Type-1.5 superconductivity in two-band systems
}

\author{Egor Babaev${}^{1,2}$ and Johan Carlstr\"om${}^{1}$}

\affiliation{
${}^1$ The Royal Institute of Technology, Stockholm, SE-10691 Sweden\\
${}^2$ Department of Physics, University of Massachusetts Amherst, MA 01003 USA}

\begin{abstract}
In the usual Ginzburg-Landau theory the critical value 
of the ratio of two fundamental length scales in the thery $\kappa_c=1/\sqrt{2}$
separates regimes of type-I and type-II superconductivity. The { latter} 
regime possess thermodynamically stable
vortex excitations which interact with each other  repulsively and {  tend }
to form vortex
lattices. It was shown in \cite{bs1}
that this dichotomy in broken in $U(1)\times U(1)$ Ginzburg-Landau models which
possess three fundamental length scales
which results in the exisrtence of a distinct phase with vortex excitations which interact attractively at large length scales
and repulsively at shorter distances. Here we briefly
review these results in particular discussing the role
of interband Josephson coupling 
and the case where only one band is superconducting 
while superconductivity in another band 
is induced by interband proximity effect. 
The report is  partially based  on E. Babaev, J. Carlstr\"om, J. M. Speight arXiv:0910.1607.
\footnote{A talk given at the ``Vortex VI" conference on 17 September 2009, Rhodes, Greece.}
\end{abstract}
\maketitle

\begin{figure}
\begin{center}
\includegraphics[width=90mm]{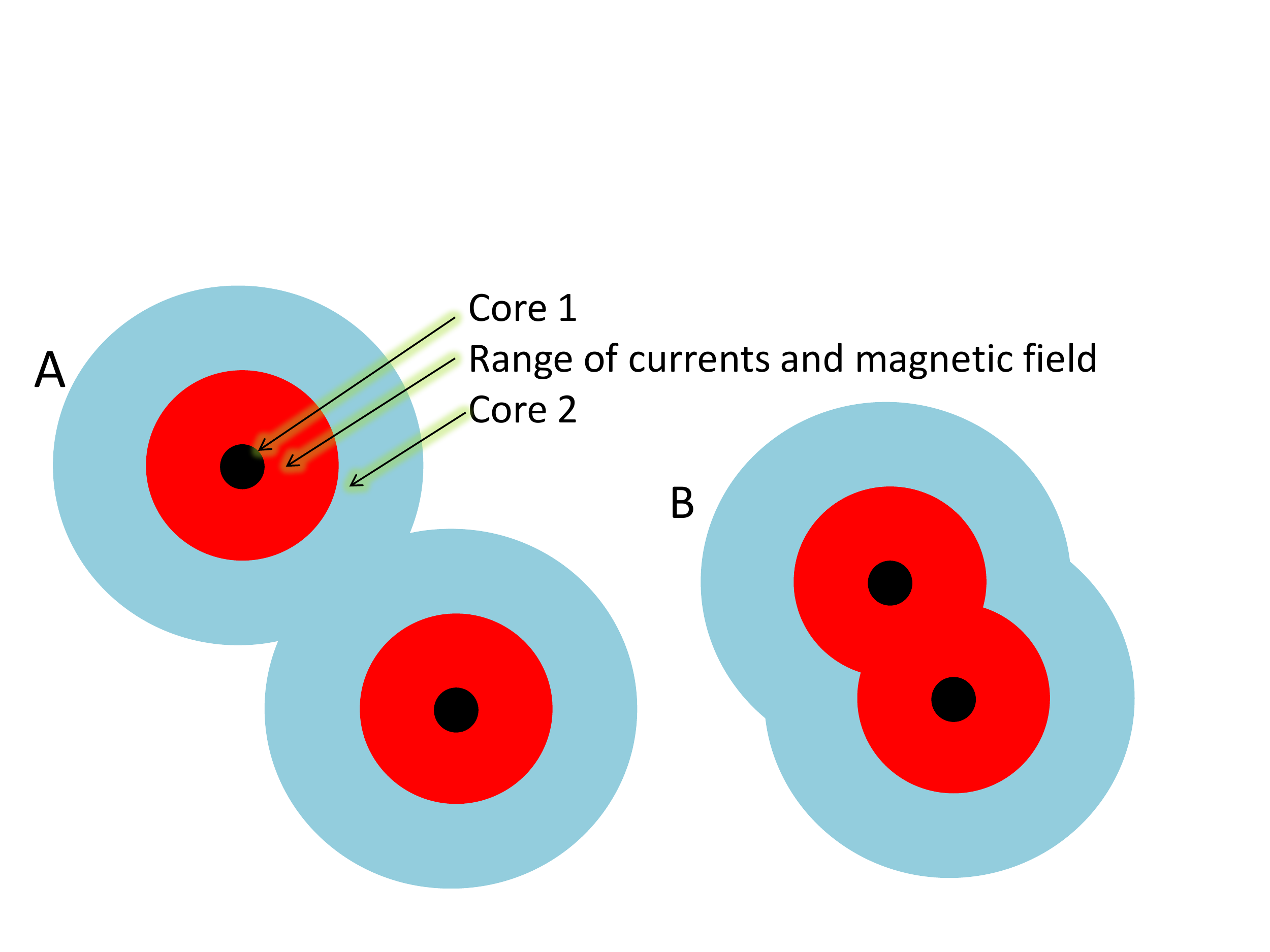}
\end{center}
\caption{A schematic illustration   of the origin of the non-monotonic interaction potential between
vortices in the superfluid mixture without intercomponent Josephson coupling 
discussed in \cite{bs1}. A: attractive interaction mediated by outer cores overlap B:
domination of the repulsive interaction mediated by currents and magnetic field. }
\label{cartoon}
\end{figure}

The textbook classifications of superconductors divide them in two classes: type-I and type-II, according 
to their behavior in an external field. Type-I superconductors expel low magnetic fields,
while elevated fields produce macroscopic normal domains in the interior of superconductor.
Type-II superconductors possess much richer magnetic response by supporting stable vortex excitations.
Lattices of these vortices form as the energetically preferred state when the applied magnetic field exceeds
a certain threshold called the lower critical magnetic field. 
This picture of type-II superconductivity
relies on the fact that interaction between co-directed vortices is purely repulsive \cite{abrikosov}.
In \cite{bs1} is was demonstrated  that in two-component superconductors, there are vortex solutions 
in a very wide parameter range 
which are on one hand thermodynamically stable, and on the other hand, possess interaction potential
which is non-monotonic: repulsive at short distances but attractive at larger distances.
The longer range attractive interaction part  originates from the fact that  in these solutions, 
the size of the core of one of the components
is the largest length scale of the problem: i.e. the core of one of the components
extends beyond the current carrying region.
In general, the precise conditions for the appearance of non-monotonic interaction are
quite complicated. However, in the simplest case the following description is 
quite accurate:
When two vortices are situated at a distance smaller than the extended core size, but larger than 
the effective magnetic field penetration length, then the vortices attract each other. At shorter distances
the interaction mediated by currents and magnetic field wins and the vortices start to repel each other.
This is schematically shown on Fig. \ref{cartoon}. 
 It should be stressed that in the one-component Ginzburg-Landau theory co-directed vortices have attractive interaction
they are thermodynamically unstable because 
the first critical magnetic field in that case is typically larger than the thermodynamical
 critical magnetic field. However it was shown that in two-component superconductors
 there is a large range of parameters where the vortices with long-range
attractive, and short-range repulsive interaction are thermodynamically stable 
(i.e. can be produced by magnetic fields with strengths smaller than 
the thermodynamical critical magnetic field  \cite{bs1}).

Indeed such a
vortex interaction, along with thermodynamic stability, should cause the system response to external field
to be entirely different from vortex states of traditional type-II Ginzburg-Landau model. Namely, the 
attraction between vortices 
should, at low fields, produce the ``semi-Meissner
state" \cite{bs1}). The implications of it include (i) formation of voids of vortex-less states, where 
there are {\it two well developed superconducting components} and (ii) 
{\it vortex clusters where one of the components
would typically dominate} because  the second component
would be suppressed (in fact significantly suppressed for a range of parameters) due to  overlapping of outer cores of the vortices.
 The ``phase separation", of this nature, which, from the point of view of the second  component resembles a mixed state of type-I superconductors
 makes this system principally different from the inhomogeneous vortex states of 
single-component superconductors where inhomogeneity can be induced by  small
corrections beyond the Ginzburg-Landau theory in regimes where  $\kappa$ is close to $1/\sqrt{2}$ see remark \cite{remark}.

The  two-band superconductor $\textrm{MgB}_2$  \cite{mgb} was regarded in many early theoretical and experimental works 
as a standard  type-II superconductor which should possess regular Abrikosov vortex lattices \cite{agt}. However, an objection
to this scenario was raised in the recent
experimental works by  Moshchalkov et  al. \cite{m1,m2} where 
a formation of  highly inhomogeneous states
was observed  with vortex clusters and vortex-less Meissner domains
strikingly similar to the   picture of 
the semi-Meissner state \cite{bs1} which results from vortices having a longer range attractive part in the interaction potential
in the two-component model.
In the ref. \cite{m1}
which was based on Bitter decoration methods and
\cite{m2} based on scanning SQUID microscopy
a statistically preferred intervortex separation was
reported.
 Moshchalkov et al. proposed that this phase separation is
an intrinsic property of $MgB_2$ and is associated with the mentioned
above three fundamental length scales in a two-component superconductor
which in that case represents a new kind of superconducting states outside
the usual type-I/type-II dichotomy. 
The term type-1.5 superconductivity was
coined for this scenario in \cite{m1}.
Let us stress that 
if there appear several fundamental length scales at the level of Ginzburg-Landau theory
such a state is indeed entirely different from
the states of single-component supercondtors. In the latter case, although a variety of different non-universal microcopic corrections
may indeed produce a weak intervortex attraction \cite{remark}, it does
not alter the classification of single-component superconductors
at the level of fundamental length scales in the Ginzburg-Landau theory.

The theory in \cite{bs1} with added intercomponent 
Josephson coupling (briefly considered below) directly applies to the case where there is fully developed 
 superconductivity in both bands. 
 However in general in a two-band superconductor, at elevated 
temperatures there  can be a regime where only one band is superconducting 
 while superconductivity in another band
{ is} induced by inter-band  proximity effect (also called inter-band Josephson effect).
In particular 
{ this }was argued to be the case in $\textrm{MgB}_2$ above a certain temperature \cite{gurevich}.

{ So it is an important generic question
 { whether} type-1.5 superconductivity is possible in the case where one of the bands 
 does not have a coherence length in the Ginzburg-Landau sense, 
 and has a non-zero density of superconducting condensate only because of the inter-band proximity effect.}

To study the essential properties of vortex physics in two-component systems
we use the following free energy density functional

\begin{eqnarray}
\label{ind_energy}
\mathcal{F}=  \frac{1}{2}\Big(|\psi_1|^2-1\Big)^2+\alpha|\psi_2|^2 +\frac{1}{2}\beta|\psi_2|^4\\\nonumber
+\frac{1}{2}|(\nabla+ ie{\bf A}) \psi_1  |^2+   \frac{1}{2}|(\nabla+ ie{\bf A}) \psi_2  |^2 \\\nonumber
-\eta|\psi_1|| \psi_2|\cos(\theta_2-\theta_1)+\frac{1}{2}   (
\nabla \times {\bf A}
)^2. 
\end{eqnarray}
The regime with $\eta=0,\alpha<0,\beta>0$ corresponds to the situation of two independent
superconducting components coupled only by vector potential, studied in \cite{bs1}.
In the case of two bands with well-developed superconductivity, the inter-band
Josephson coupling  $\eta\ne 0$ works against the type-1.5 regimes.

\begin{figure}
\begin{center}
\includegraphics[width=78mm]{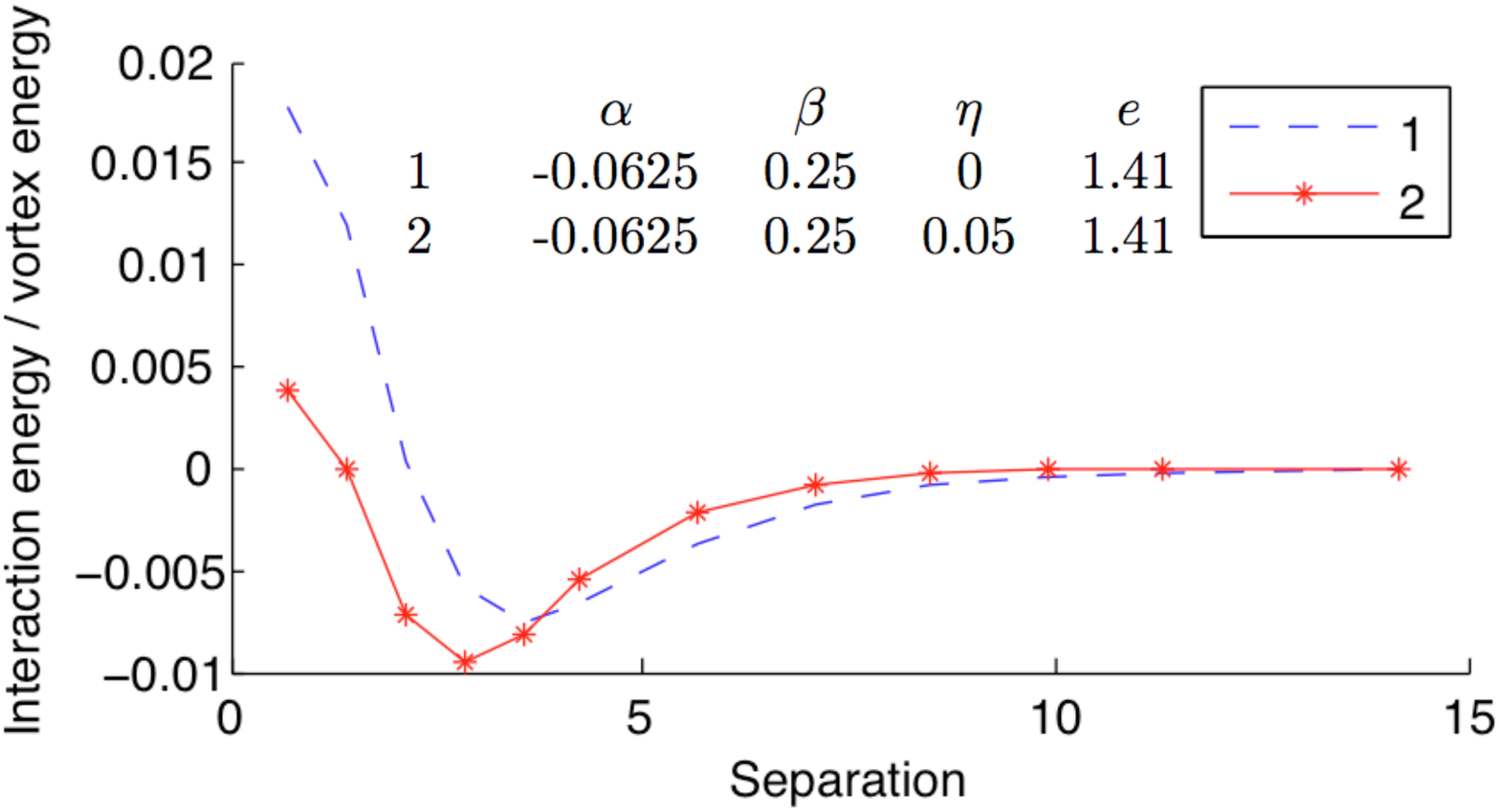}
\end{center}
\caption{Intervortex interaction energy in a system with two active bands. In the first case, the Josephson coupling is zero, and the ground state
 densities of the condensates are $1$ and $0.25$. In the 
second case, nonzero but moderate Josephson coupling $\eta=0.05$ decreases the range of the attractive part of the interaction potential
but at the same time it increases the ground state densities to approximately $1$ and $0.4$, yielding a slightly deeper minimum. }
\label{twoband}
\end{figure}

We studied numerically 
the effect of the 
Josephson coupling on the
vortex-vortex interaction energy in a system with two superconducting bands
(i.e. $\eta\ne 0,\alpha<0,\beta>0$). The results of numerical calcualions
of the intervortex interaction energy in the model (\ref{ind_energy}) are
shown on   Fig. \ref{twoband}.  In the first curve $\eta=0$ and the condensates interact only through the shared vector potential,
the parameters $\alpha,\beta,e$ were choosen to yield a disparity of coherence lengths and penetration depth to produce
 a   type-1.5 regime. Adding a moderate Josephson coupling  $\eta=0.05$ increases ground state densities
of the condensates, and decreases penetration length (which depends on superfluid densities in both bands and thus on $\eta$)
 which results in fact in a {\it deeper} minimum of the interaction potential. 
However this coupling  decreases the disparity of the recovery rates of the 
condensates, resulting in a decreased range of the attractive interaction.
Even though  a sufficiently strong Josephson coupling in the GL model
can eliminate type-1.5 behavior, this example shows that 
the type-1.5 behavior survices even in case of 
a rather substantial interband Josephson coupling.
 Similarly type-1.5 regime exists also in the 
 presence of mixed gradient terms \cite{bcs2}.

Consider now the the case of nonzero Josephson coupling $\eta \ne 0$
but with  one of the bands being {\it above} its critical temperature \cite{bcs}.
In that case the effective potential for $\psi_2$ has  {\it only positive} coefficients $ \alpha,\beta >0$.
Thus the second band has a nonzero density of Cooper pairs only because of inter-band tunneling represented by the term $-\eta|\psi_1|| \psi_2|\cos(\theta_2-\theta_1)$.
This term also locks phases $\theta_1=\theta_2$. So in the following, we consider only solutions with the winding 
in the total ``locked" phase. These vortices have finite energy and carry one flux quantum. If there is a phase winding only in one phase,
one gets a Josephson vortex with linearly diverging energy \cite{prl02} (which cannot be produced by external field under usual circumstances). 

Since the phases in this regime are locked to equal values
which minimizes the Josephson term $\theta_1=\theta_2=\phi$,
our effective model becomes  
\begin{eqnarray}
\label{ind_energy2}
\mathcal{F}=  \frac{1}{2}\Big(|\psi_1|^2-1\Big)^2+\alpha|\psi_2|^2 +\frac{1}{2}\beta|\psi_2|^4\\\nonumber
+\frac{1}{2}|(\nabla+ ie{\bf A}) \psi_1  |^2+   \frac{1}{2}|(\nabla+ ie{\bf A}) \psi_2  |^2 \\\nonumber
-\eta|\psi_1|| \psi_2|+\frac{1}{2}   (
\nabla \times {\bf A}
)^2. 
\end{eqnarray}

We present accurate numerical solutions for one-quanta vortices (i.e. with the phase winding $\Delta \phi =2\pi$) 
and vortex-vortex interaction in the model (\ref{ind_energy2}) 
(for other details including analytical theory see \cite{bcs}).
The numerical solutions were obtained using a local relaxation method. 
A two-vortex configuration 
is initially generated fixing only the 
positions of the vortex cores and phase windings.
Then this multiple vortex configuration is relaxed with respect to all the other degrees of freedom
in the system, thus producing highly accurate 
solutions of the Ginzburg-Landau equations of motion with given phase windings and vortex separation. 
The procedure is repeated for a different vortex separation yielding a highly accurate vortex interaction potential.


\begin{figure}
\begin{center}
\includegraphics[width=90mm]{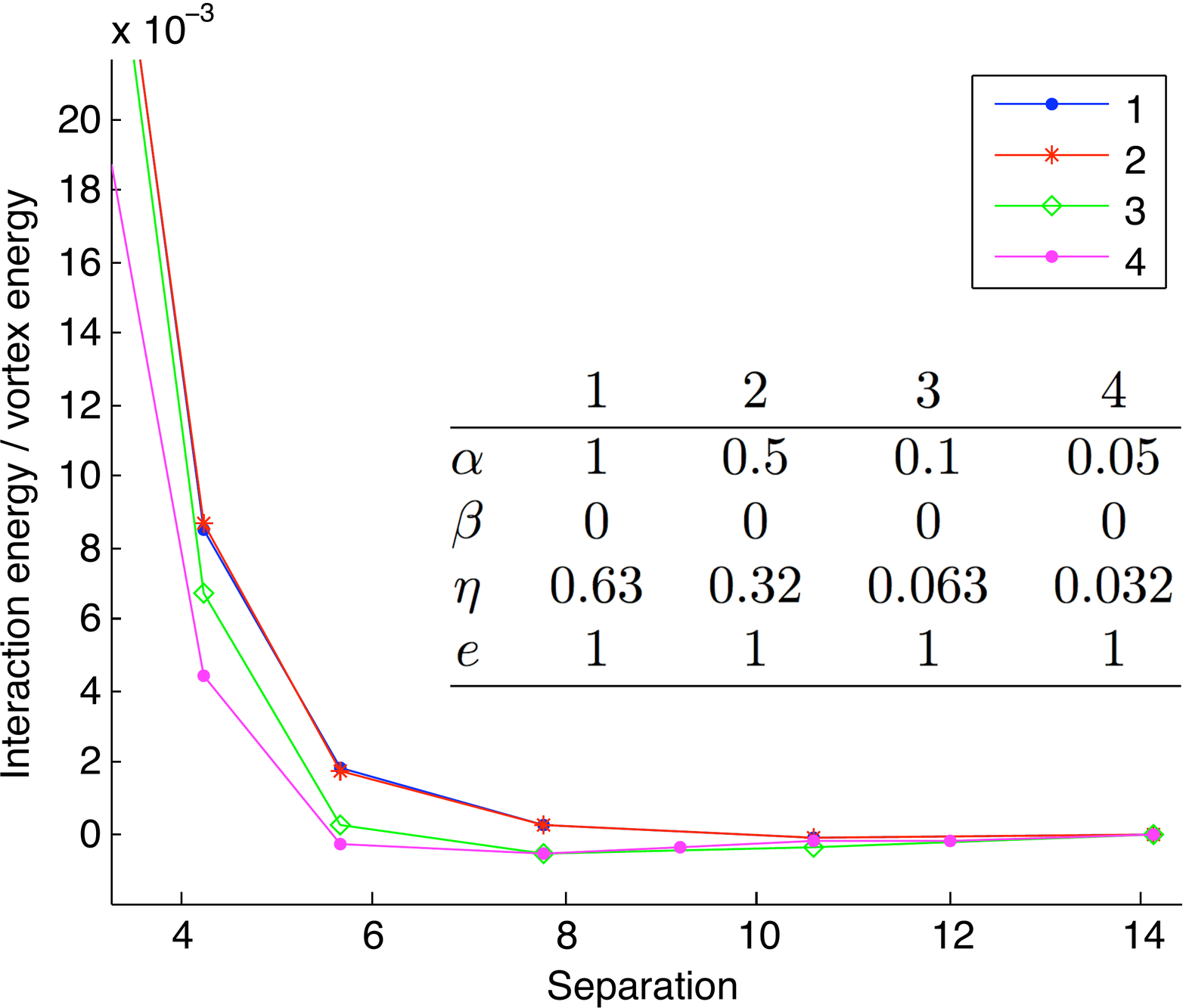}
\end{center}
\caption{Intervortex interaction energy for a density ratio of $0.1$.  }
\label{s1}
\end{figure}

\begin{figure}
\begin{center}
\includegraphics[width=86mm]{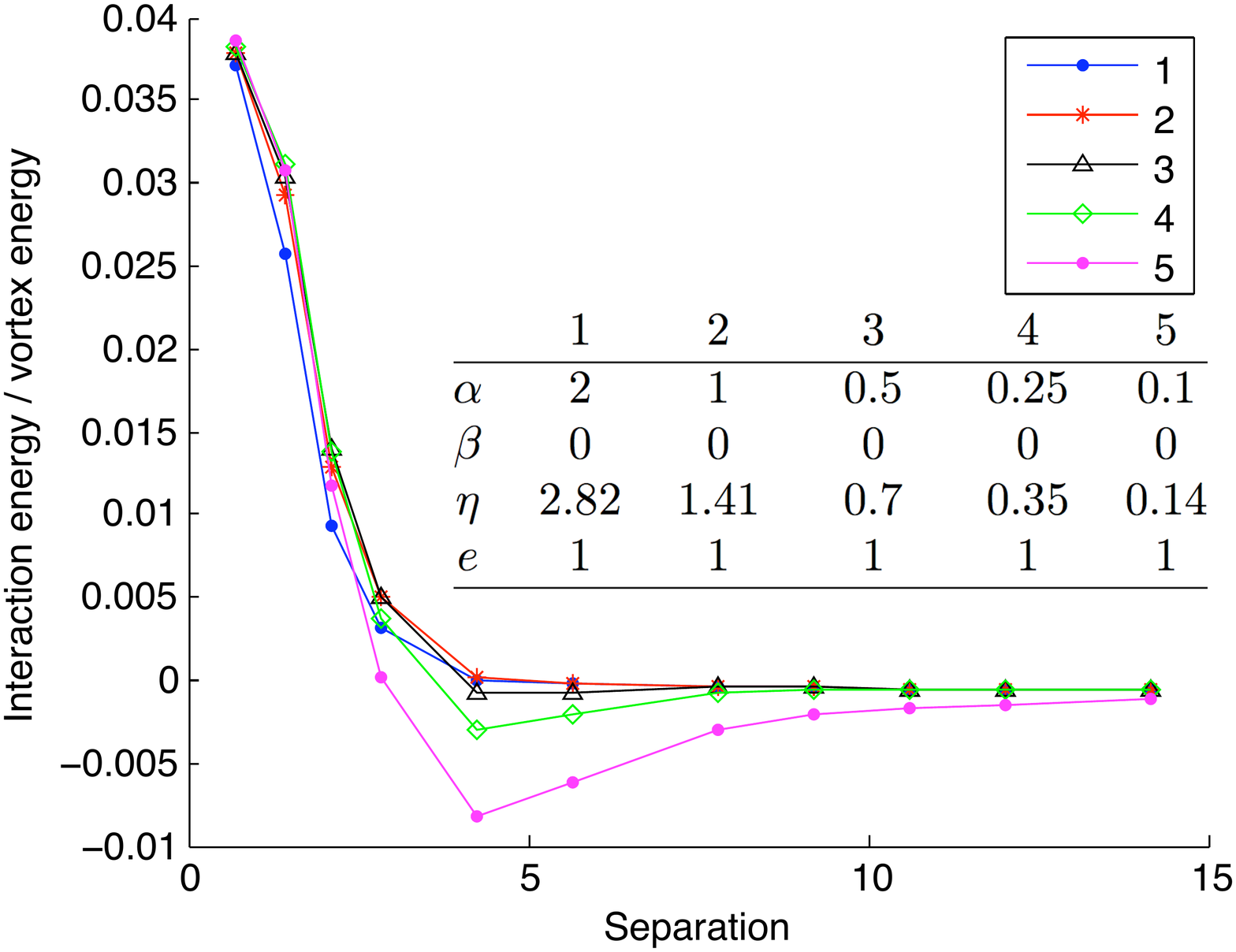}
\end{center}
\caption{Intervortex interaction energies for density ratio of $0.5$.  }
\label{s2}
\end{figure}

\begin{figure}
\begin{center}
\includegraphics[width=86mm]{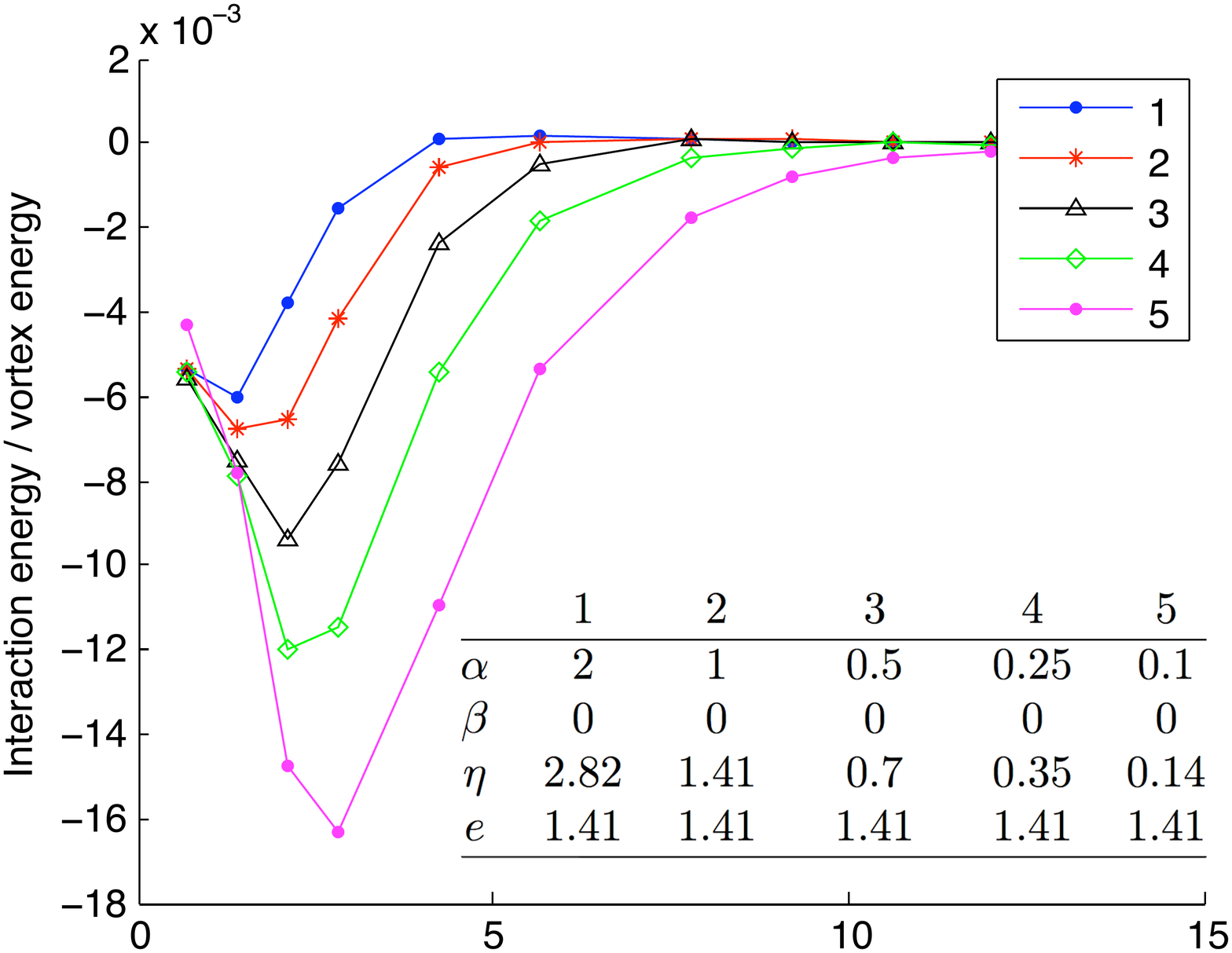}
\end{center}
\caption{Intervortex interaction energies at a density ratio of $0.5$ and an increased charge of $e=1.41$.}
\label{s3}
\end{figure}

First lets consider the regime where the fourth order term in $|\psi_2|$ can be neglected.
In this case we conducted simulations with the density ratios $|\psi_2|^2/|\psi_1|^2$ being $0.1$ and $0.5$ \cite{bcs}.
The numerical results  are presented in Figs. \ref{s1}-\ref{s3}. 
The computed  interaction energy is given in units of $2E_v$ where $E_v$ is the energy of a single vortex. 

In the first case with the density ratio $0.1$,  we find that 
in general, the recovery lengths of the condensates can be quite different,
even though one  of the bands has proximity-induced superconductivity.
We find that as a consequence of the disparity in the recovery lengths, the system 
 crosses over from the Type-II to the Type-1.5 regime when $\alpha$ and  $\eta$ are sufficiently 
 small
 (Fig. \ref{s1}).  The low density of condensate in the  band with proximity-induced superconductivity 
means that the attractive part of the interaction is weak. In the curves 3 and 4, 
we find  a slight long range attraction, yielding a minimum energy at a separation of approximately 8. 
The curves 3 and 4 correspond to the smallest values of $\alpha$ and $\eta$, yielding quite large cores in the  band with proximity induced superconductivity.

In the second case (Fig. \ref{s2}), the density ratio is $0.5$. The vortex-vortex binding energy is now much larger, and the minimum energy occurs at a smaller separation. Long range attraction occur in curves 3-5 with a maximum $\alpha$ of $0.5$, in contrast to   $\alpha \approx 0.1$ in the previous case. 

In the third case (Fig. \ref{s3}), the charge has been increased by a factor $\sqrt{2}$. The resulting shorter penetration length
decreases the magnetic repulsion between vortices. Observe that now the energy of an axially symmetric vortex solution 
with two flux quanta is smaller than the energy of two infinitely separated one-quanta vortices, nonetheless the axially-symmetric 
two-quanta vortex is not stable since the minimum energy corresponds to a nonzero vortex separation.

\begin{figure}
\begin{center}
\includegraphics[width=70mm]{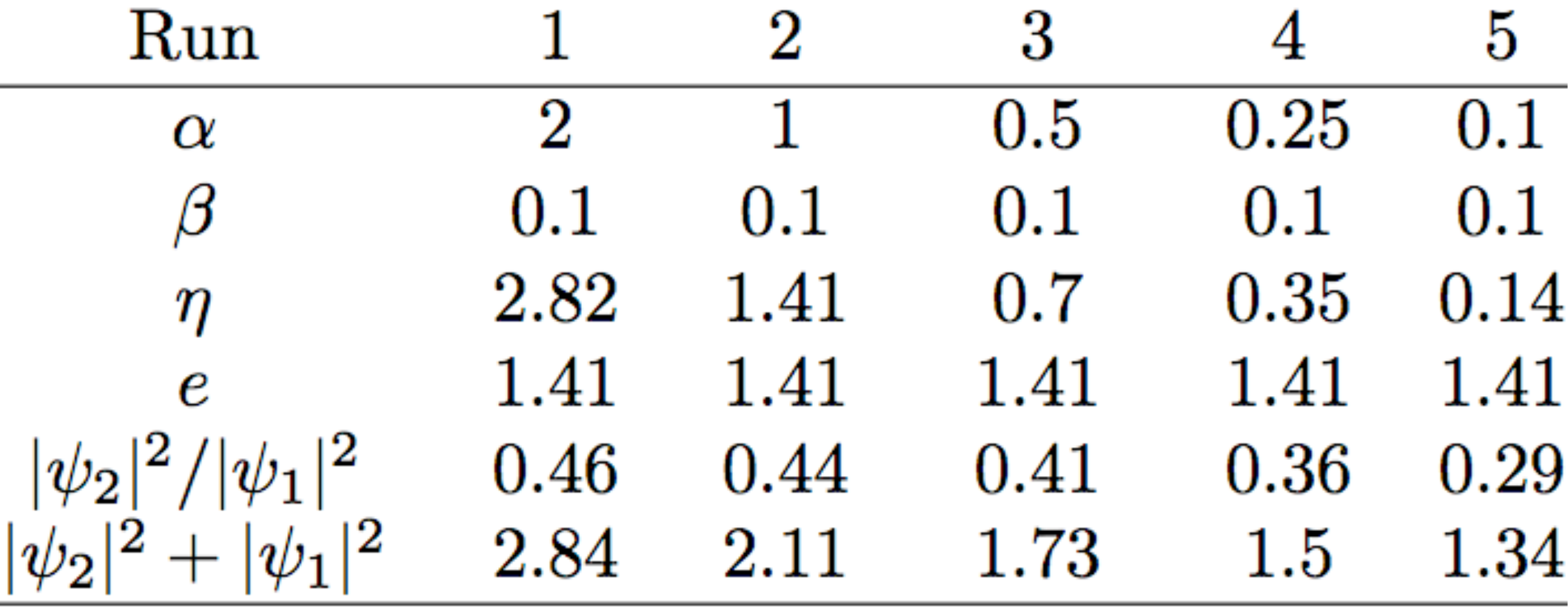}
\includegraphics[width=90mm]{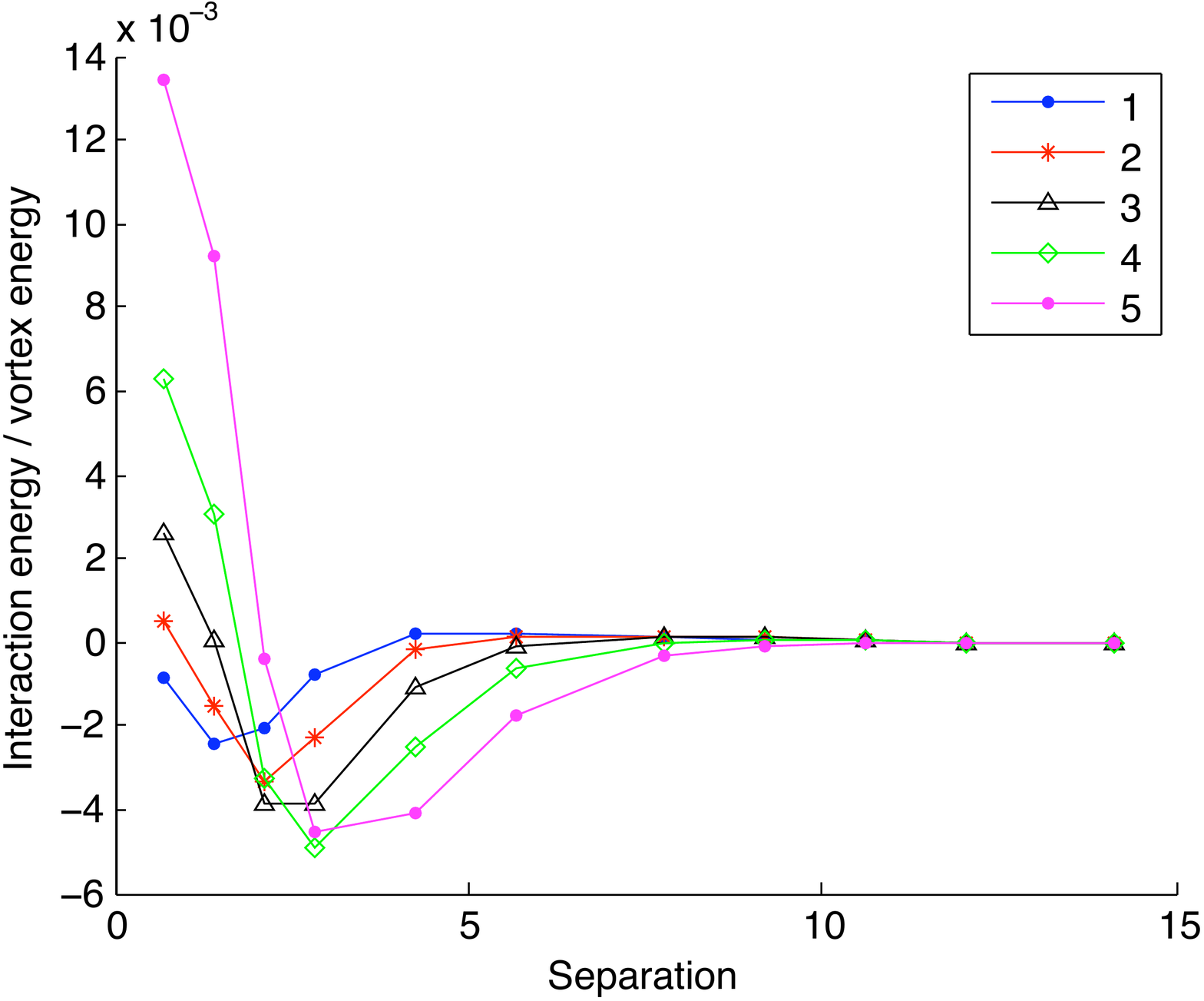}
\end{center}
\caption{Intervortex interaction energy. Model parameters are given in the  inset. Observe that the density ratios are different
for different curves.}
\label{s4}
\end{figure}
\begin{figure}
\begin{center}
\includegraphics[width=82mm]{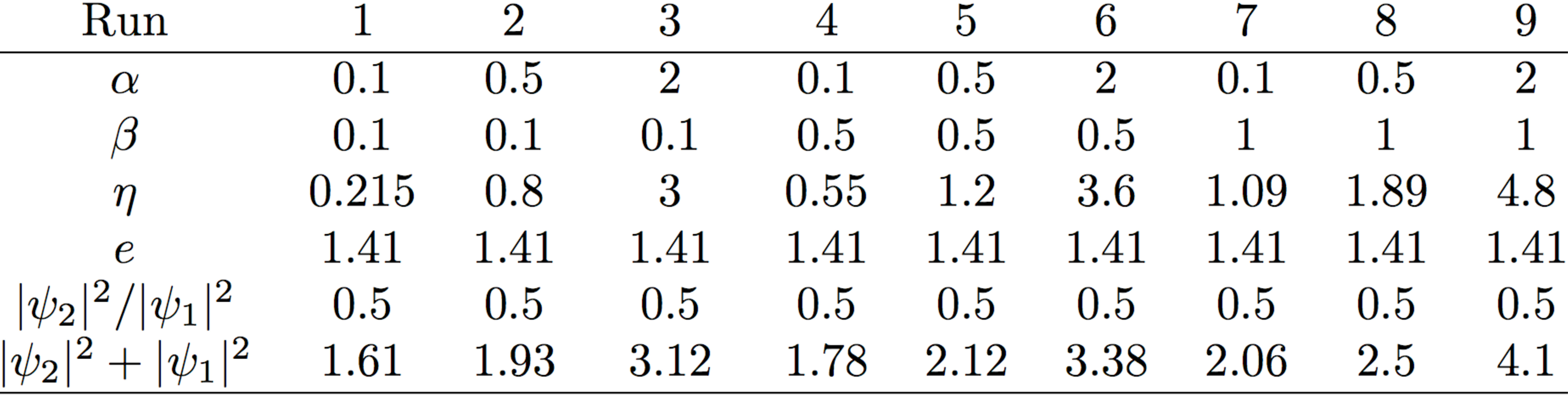}
\includegraphics[width=90mm]{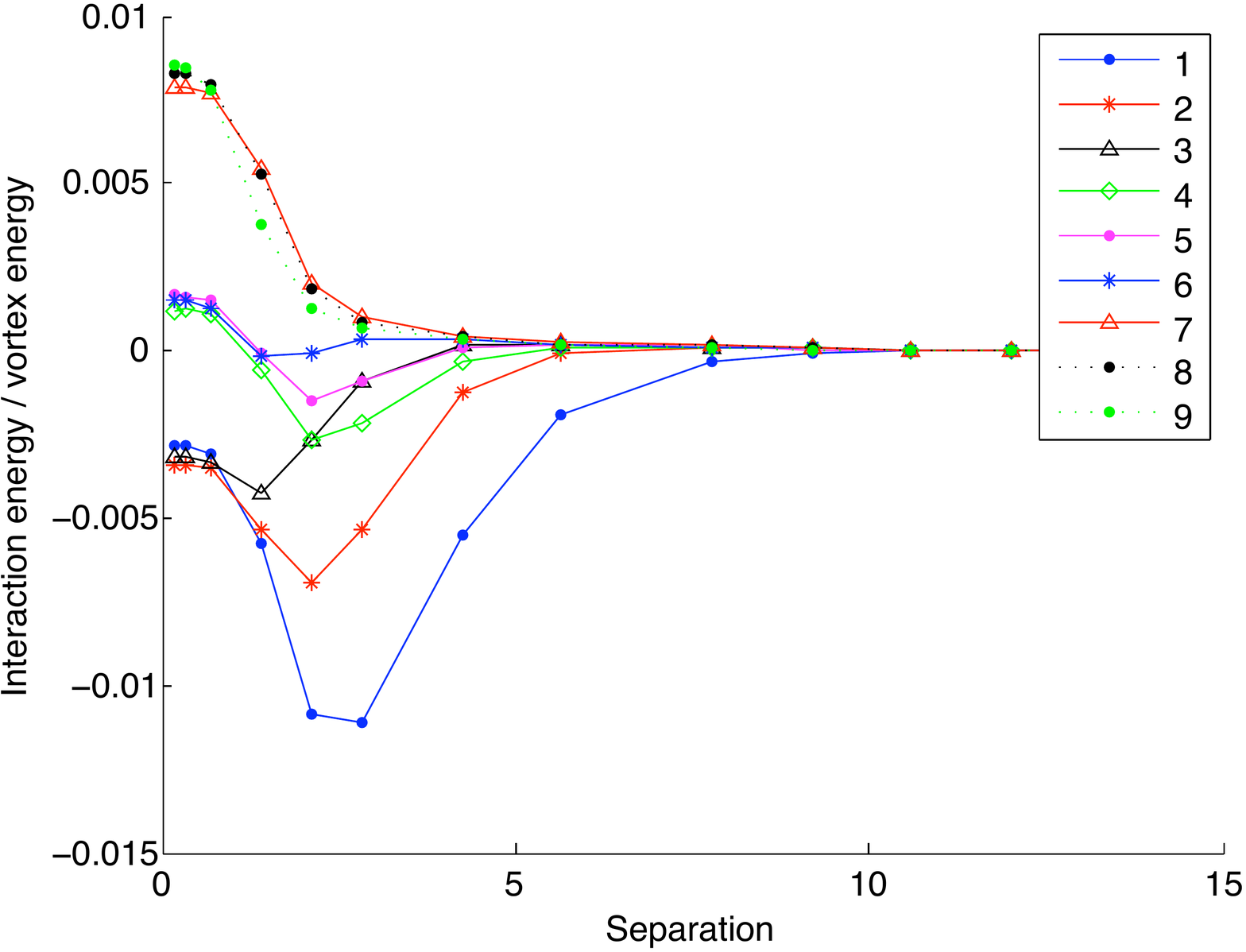}
\end{center}
\caption{Intervortex interaction energy in different regimes described in the inset.}
\label{s5}
\end{figure}
\begin{figure}
\begin{center}
\includegraphics[width=90mm]{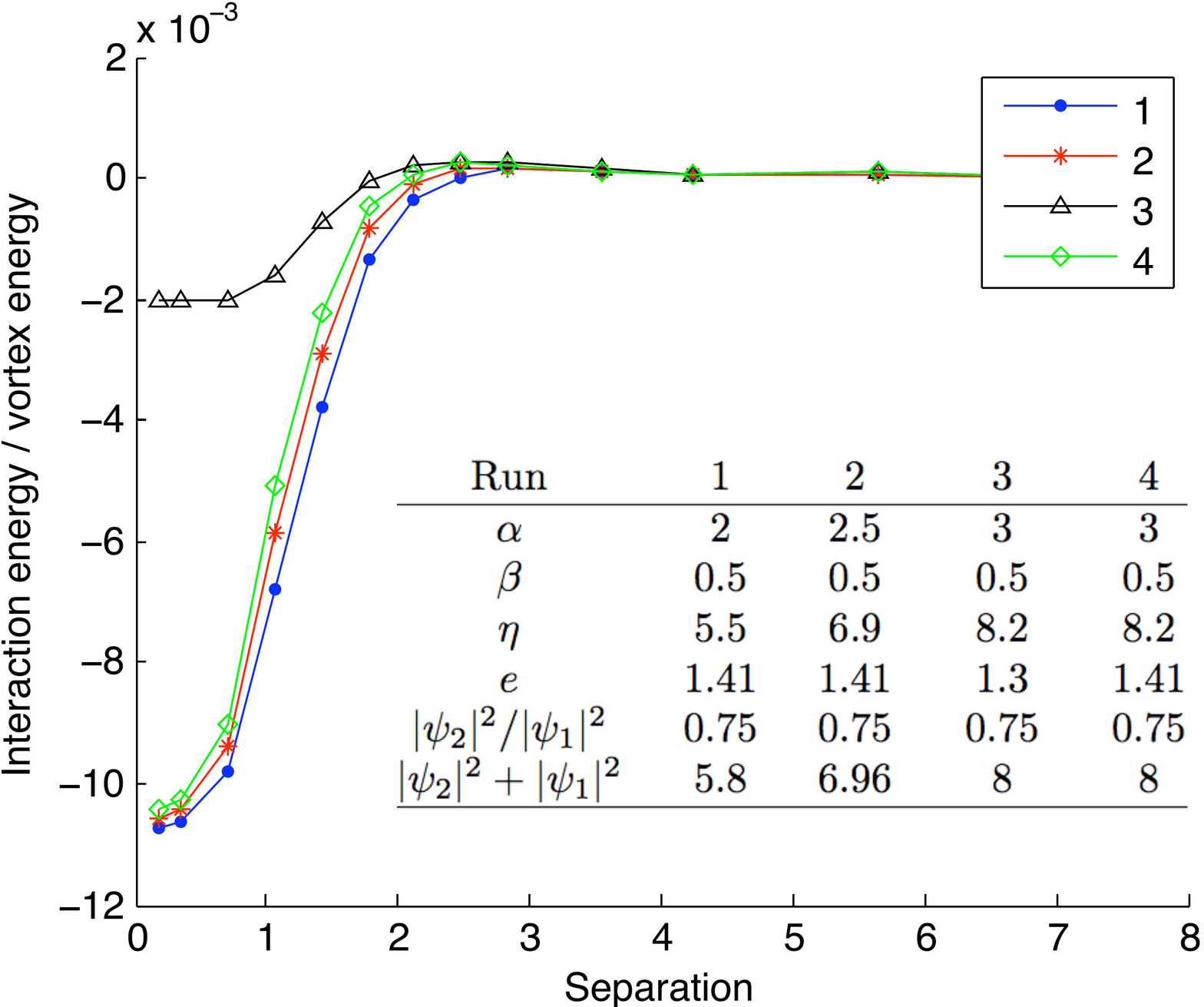}
\end{center}
\caption{Intervortex interaction energy in different regimes described in the inset. }
\label{s6}
\end{figure}
Figure \ref{s4} shows the effect of the addition of a fourth order term  with $\beta=0.1$ in the free energy 
of the band with Josephson-induced superconductivity. 
 This image is further reinforced in Fig. \ref{s5} showing 
pronounced non-monotonic interaction and thus type-1.5 superconductivity in this system. 
\begin{figure}
\begin{center}
\includegraphics[width=85mm]{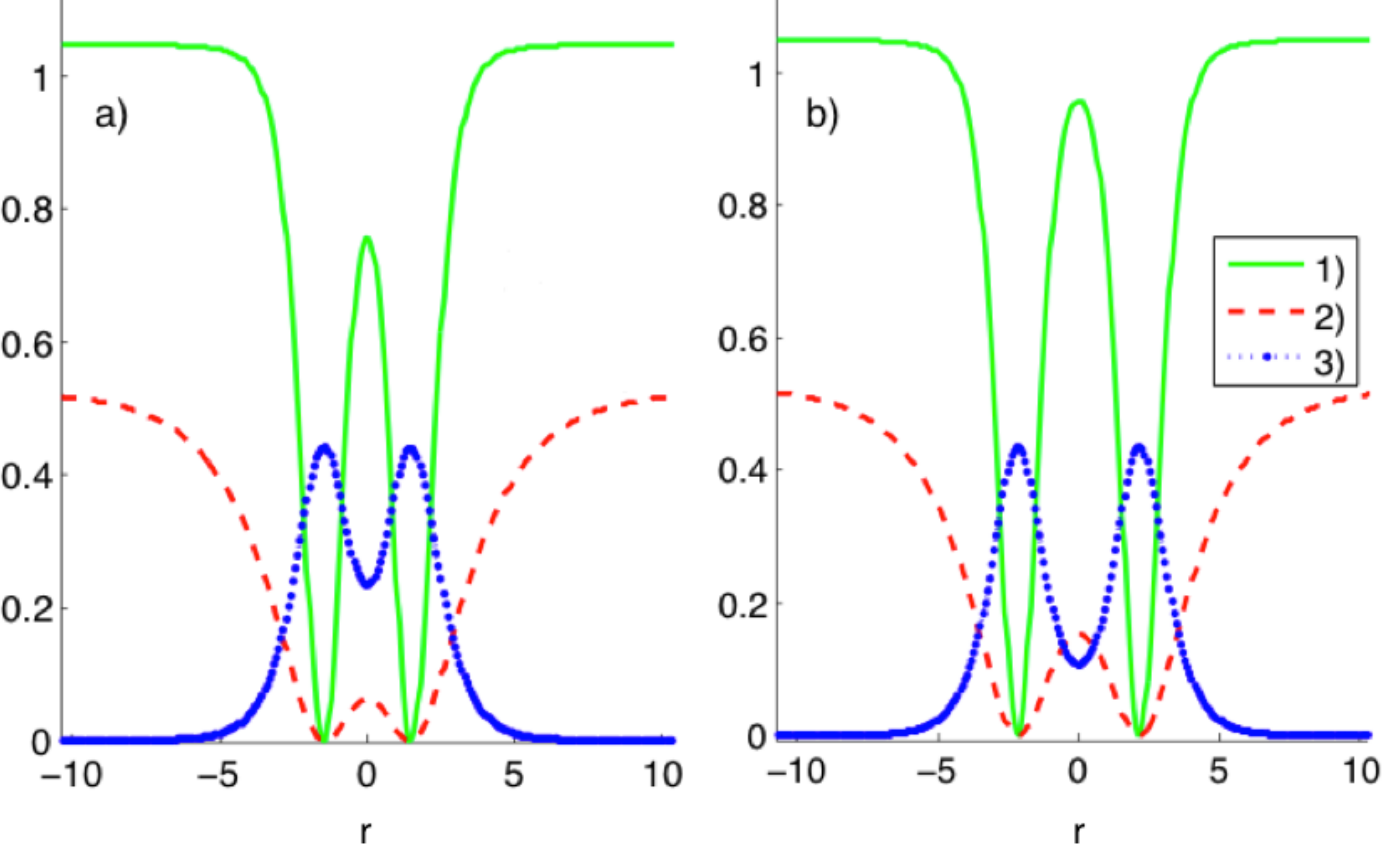}
\end{center}
\caption{Cross section of interacting vortices for the case $\alpha=0.1$, $\beta=0$, $\eta=0.14$ and $e=1$ (the fifth curve in fig \ref{s2}). 
Curves ``1" and ``2" show the behaviour of $|\psi_{1,2}|$ and the curve ``3" represents the magnetic field.
The right image displays the system at 
vortex separation of $\approx 4.2$ (energy minimum),  the left image displays the system at a separation of $\approx 2.8$ }
\label{cross1}
\end{figure}
\begin{figure}
\begin{center}
\includegraphics[width=85mm]{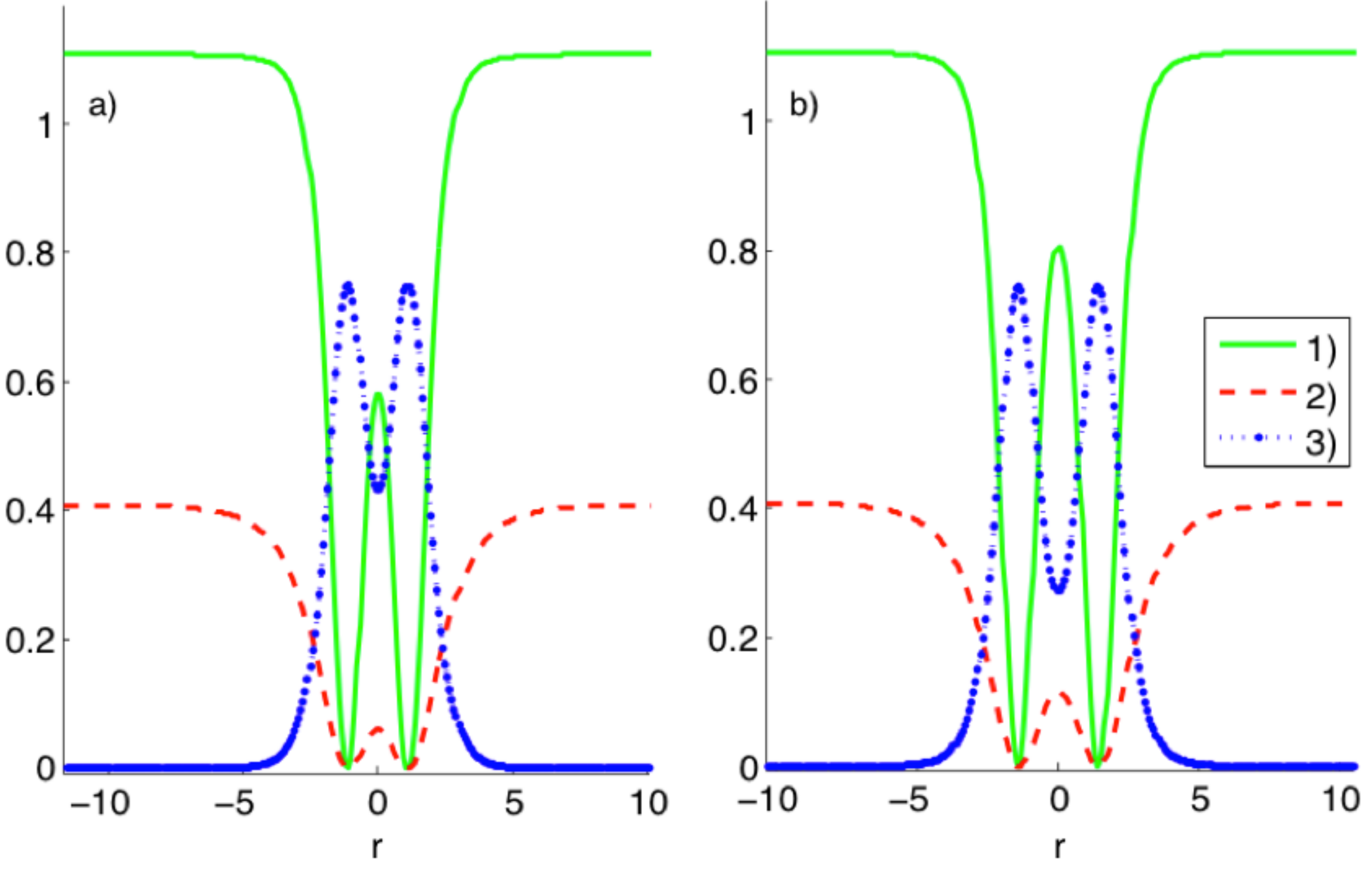}
\end{center}
\caption{Cross section of interacting vortices for the case $\alpha=0.25$, $\beta=0.1$, $\eta=0.35$ and $e=1.41$ (the fourth curve in fig \ref{s4}). 
Curves ``1" and ``2" show the behaviour of $|\psi_{1,2}|$ and the curve ``3" represents the magnetic field.
The right image displays the system at a separation of $\approx 2.8$ (energy minimum),  the left image displays the system at a separation of $\approx 2.1$ }
\label{cross2}
\end{figure}
Figure \ref{s6} shows a system with larger density ratio than the previous systems. The increased 
condensate density, especially in the band 
with Josephson-induced condensate provides a dominating attractive interaction potential that pushes the system into the Type-I regime.

The figures \ref{cross1}-\ref{cross2} show cross-sections of vortices in two  cases exhibiting  Type-1.5 superconductivity.  The first case is the fifth curve of Fig. \ref{s2}. The right image correspond to the energy minimum. Here, the cores overlap is significant in the induced band, but almost nonexistent in the active band. There is a moderate overlap of magnetic fields. Decreasing the separation produces slightly more cores overlap in the band
with induced superconductivity, but the condensation energy gained is more than compensated  by the increasing magnetic and current-current interaction-driven repulsion, resulting in increased total energy.  
In the second case, corresponding to the fourth curve in Fig. \ref{s4}, the charge is larger, resulting in a more sharply peaked magnetic field. The minimum energy does in this case occur at a smaller separation (2.8 instead of 4.2). The  overlap of the cores in the main band, as well as in the magnetic field/current carrying regions is larger in this case. 
Observe that the increase in $\alpha$ and $\eta$ clearly results in a faster recovery of the condensate density in the band with proximity effect induced superconductivity.

In conclusion, type-1.5 superconductivity is a magnetic response 
possible in multicomponent systems because of the existence of several
fundamental length scales associated with the masses of the fields
which is distinct from the type-I/type-II dichotomy found in usual single-component Ginzburg-Landau model.
Here we discuss that this kind of superconductivity may be 
present for a rather large range of parameters
in two-band systems becasue it persists in the presence of
intercomponent Josephson coupling and even can take place in the case where only one
of the bands has true superconductivity while superconductivity in the other band is induced by interband proximity effect \cite{bcs}.
\\
{\bf Acknowledgements}

We thank J. M. Speight for  collaboration on this project \cite{bcs}. We also thank Victor Moshchalkov, Alex Gurevich 
and Mats Wallin for discussions.
The work is supported by Swedish Research Council and by the Knut and Alice Wallenberg Foundation.


\begin{thebibliography}{99}
\bibitem{remark}
The single component Ginzburg-Landau theory 
is parameterized by a ratio $\kappa$
of the fundamental
length scales in the theory: the penetration length $\lambda$ 
and the coherence length $\xi$.
The critical value of the Ginzburg-Landau parameter
$\kappa_c=1/\sqrt{2}$ separates the regimes of repulsively interacting 
thermodynamically stable vortices ($\kappa_c>1/\sqrt{2}$) and attractively interacting thermodynamically
unstable vortex solutions in type-I regime ($\kappa_c<1/\sqrt{2}$). Precisely at $\kappa_c=1/\sqrt{2}$
in the Ginzburg-Landau theory vortices do not interact. Note that,
strictly speaking, the fundamental-length-scales-based
type-I/type-II classification of superconductors which yields
noninteracting vortices at the critical value of $\kappa$
is a property of a description at the level of Ginzburg-Landau
theory in the sense that more complexity can arise as
a consequence of inhomogeneity, boundary effect and indeed
at some more microscopic level one can identify a variety
of small effects which would introduce small corrections
to the universal GL form of vortex interaction.
This is quite relevant for known single-component superconductors where is $\kappa$ very close
to $1/\sqrt{2}$,  because in that regime vortex
interaction is very weak and the effects beyond the Ginzburg-Landau theory
are more pronounced and can in principle produce a weak intervortex
attraction \cite{brand}. Here however we are interested only
in a fundamental classification of magnetic responce multicomponent superconductors,
that is, a classification based
on the fundamental length scales associated with 
fields in a Ginzburg-Landau theory and we do not consider
small corrections arising from  specific microscopic physics 
or higher order corrections. 
 The effects  discussed in the 
present work and \cite{bs1,bcs} are therefore entirely different from the
attractive interaction due to non-GL effects in superconductors with
  $\kappa_c \approx 1/\sqrt{2}$ \cite{brand}. For early experiment 
 on superconductors with $\kappa$ very close to $1/\sqrt{2}$ see \cite{krag}.
\end{thebibliography}
\end{document}